\documentclass[twoside]{LCWS11}
\usepackage[latin1]{inputenc}
\usepackage[dvips]{graphicx,epsfig,color}
\usepackage{wrapfig,rotating}
\usepackage{amssymb,amsmath,array}
\usepackage{cite}

\begin{document}
\title{
Higgs at ILC in Universal Extra Dimensions\\
in Light of Recent LHC Data} 
\author{Takuya Kakuda$^1$, Kenji Nishiwaki$^2$, Kin-ya Oda$^3$, Naoya Okuda$^3$, and Ryoutaro Watanabe$^3$
\vspace{.3cm}\\
1- Department of Physics, Niigata University\\
Niigata 950-2181, Japan
\vspace{.1cm}\\
2- Harish-Chandra Research Institute\\
Chhatnag Road, Jhusi, Allahabad 211 019, India
\vspace{.1cm}\\
3- Department of Physics, Osaka University \\
Osaka 560-0043, Japan\\
}

\maketitle

\begin{abstract}
We present bounds on all the known universal extra dimension models from the latest { Higgs search data at} the Large Hadron Collider, taking into account the Kaluza-Klein (KK) loop effects {on} the dominant gluon-fusion production and {on} the diphoton/digluon decay. The lower bound on the KK scale is {from} 500\,GeV to {1}\,TeV depending on the model. We find that the Higgs production cross section { with subsequent diphoton decay} can be enhanced by a factor 1.5 within this experimental bound, with little dependence on the Higgs mass in between 115\,GeV and 130\,GeV. We {also} show that {in such a} case the Higgs decay branching ratio into a diphoton final state can be suppressed by a factor 80\%, which is marginally observable at a high energy/luminosity option at the International Linear Collider. The Higgs production cross section at a photon-photon collider can also be suppressed by a similar factor 90\%{, being well within the expected experimental reach}.
\end{abstract}

\section{Introduction}
Higgs field is the last missing and the most important piece of the Standard Model (SM) of elementary particles and interactions. Last year the Large Hadron Collider (LHC) made a great achievement in Higgs searches. Now the SM Higgs mass is highly constrained within a low mass range $115.5\,\text{GeV}<M_H<127\,\text{GeV}$ or else is pushed up to a high mass region $M_H>600\,\text{GeV}$ at the 95\% CL~\cite{ATLAS_combined,CMS_combined}.

In particular the ATLAS experiment has observed an excess of events close to $M_H=126\,\text{GeV}$ with a local significance $3.6\,\sigma$ above the expected SM background without Higgs, though it becomes less significant $2.3\,\sigma$ after taking into account the Look-Elsewhere Effect (LEE)~\cite{ATLAS_combined}. On the other hand, the CMS experiment has observed the largest excess at 124\,GeV with a local significance $3.1\,\sigma$ but reduces to $1.5\,\sigma$ after taking the LEE into account over 110--600\,GeV~\cite{CMS_combined}. Note that the peak at ATLAS is close to the CMS exclusion limit 127\,GeV, but that the CMS local significance at 126\,GeV is still $\sim2\,\sigma$~\cite{CMS_combined}. These peaks at ATLAS and CMS are dominated by diphoton signals.\footnote{
 Our analyses and statements hereof are based on the results shown in the preliminary version presented on the web in Refs.~\cite{ATLAS_combined,CMS_combined}.
}

An interesting observation is that the best fit value of the diphoton cross section is enhanced from that of SM by factor $\sim1.7$ and 2 for the peaks at $M_H=124\,\text{GeV}$ (CMS~\cite{CMS_combined}) and 126\,GeV (ATLAS~\cite{ATLAS_combined}), respectively. For the latter, the enhancement needed for the {total} Higgs production cross section is  $\sim1.5$ after taking into account all the related decay channels (with the branching ratios being assumed to be the same as in the SM): $H\to\gamma\gamma$, $H\to ZZ\to llll$ and $H\to WW\to l\nu l\nu$~\cite{ATLAS_combined}.

{ The Universal Extra Dimension (UED) models assume that} all the SM fields propagate in the bulk of the compactified extra dimension(s). Currently known UED models utilize compactifications
	on { a} one-dimensional orbifold $S^1/Z_2$ (mUED), 
	on two-dimensional orbifolds based on torus $T^2/Z_{4}$ (T2Z4), $T^2/(Z_2\times Z_2')$ (T2Z2Z2), $T^2/Z_2$ (T2Z2), $RP^2$ (RP2), 
	on { a} two-sphere based orbifold $S^2/Z_2$ (S2Z2), and 
	on two-dimensional manifolds, the projective sphere (PS) and the sphere $S^2$ (S2); See~\cite{Nishiwaki:2011gk,Nishiwaki:2011gm} for references. 

{ We can list} {two} virtues of the UED models { (see e.g.~\cite{Nishiwaki:2011gm} for references)}. {First,} due to the compactification, there appears a tower of Kaluza-Klein (KK) modes for each SM degree of freedom{; Among} these KK modes, the Lightest KK Particle (LKP) { is} stable due to a symmetry of the compactified space and hence { becomes} a good candidate for the dark matter.
Second virtue is the explanation of the number of generations to be three when there are two extra dimensions { in order to cancel the global gauge anomaly in six dimensions}. 

{Further, the UED models allow a heavy Higgs.}
If the light Higgs is excluded in the forthcoming LHC running and hence the Higgs turns out to be heavy in the region $M_H>600\,\text{GeV}$, the SM { with such a heavy Higgs} is inconsistent to the current electroweak precision data. {In UED model} the KK top loop corrections { may cure}  this discrepancy. { However} in this work, we pursue the case for light Higgs mass and give a possible explanation for the above mentioned enhancement of the Higgs production cross section.

\begin{figure}
\includegraphics[width=0.24\columnwidth]{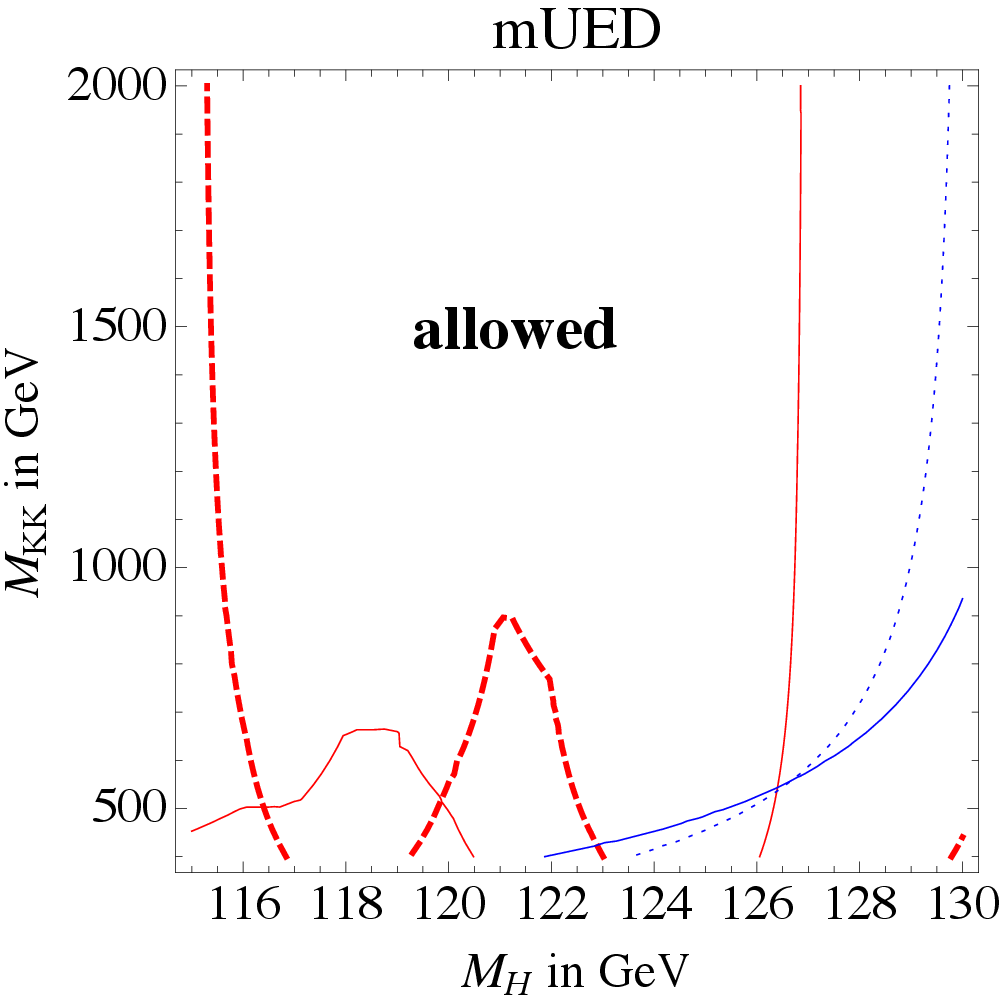}
\includegraphics[width=0.24\columnwidth]{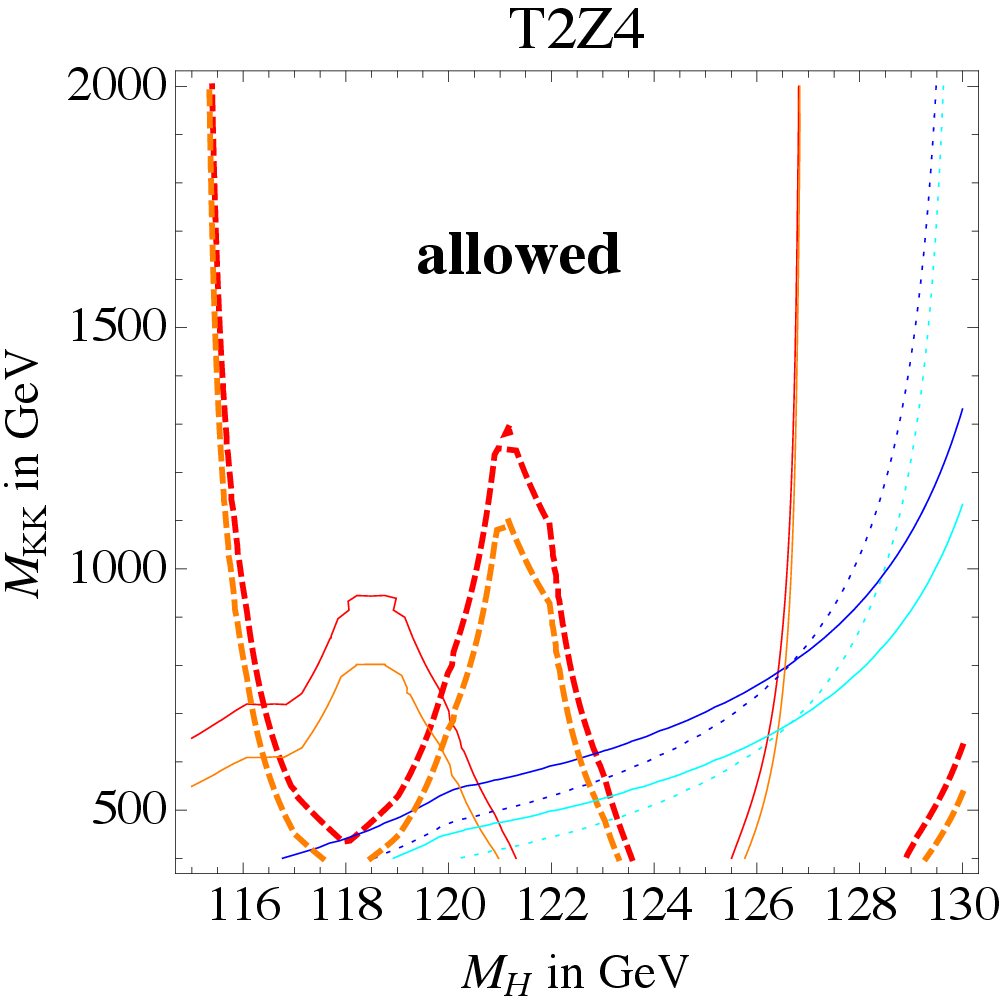}
\includegraphics[width=0.24\columnwidth]{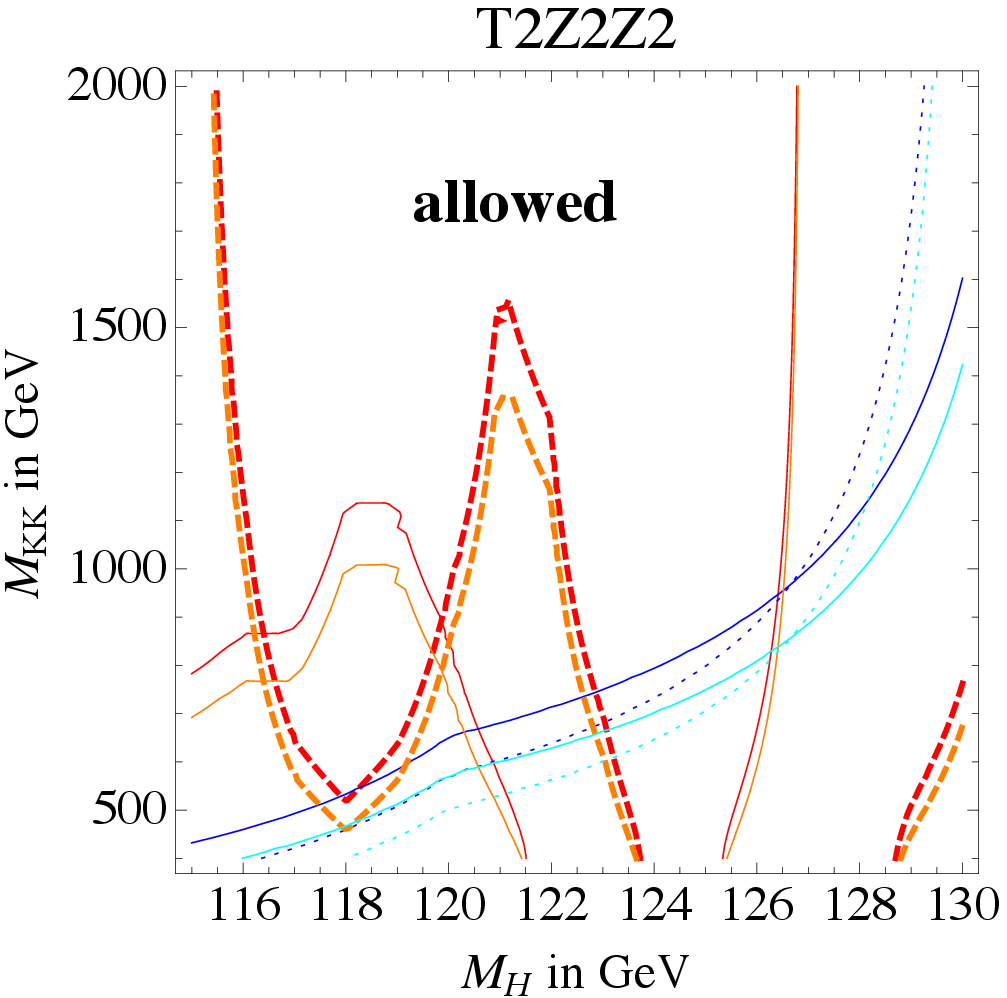}
\includegraphics[width=0.24\columnwidth]{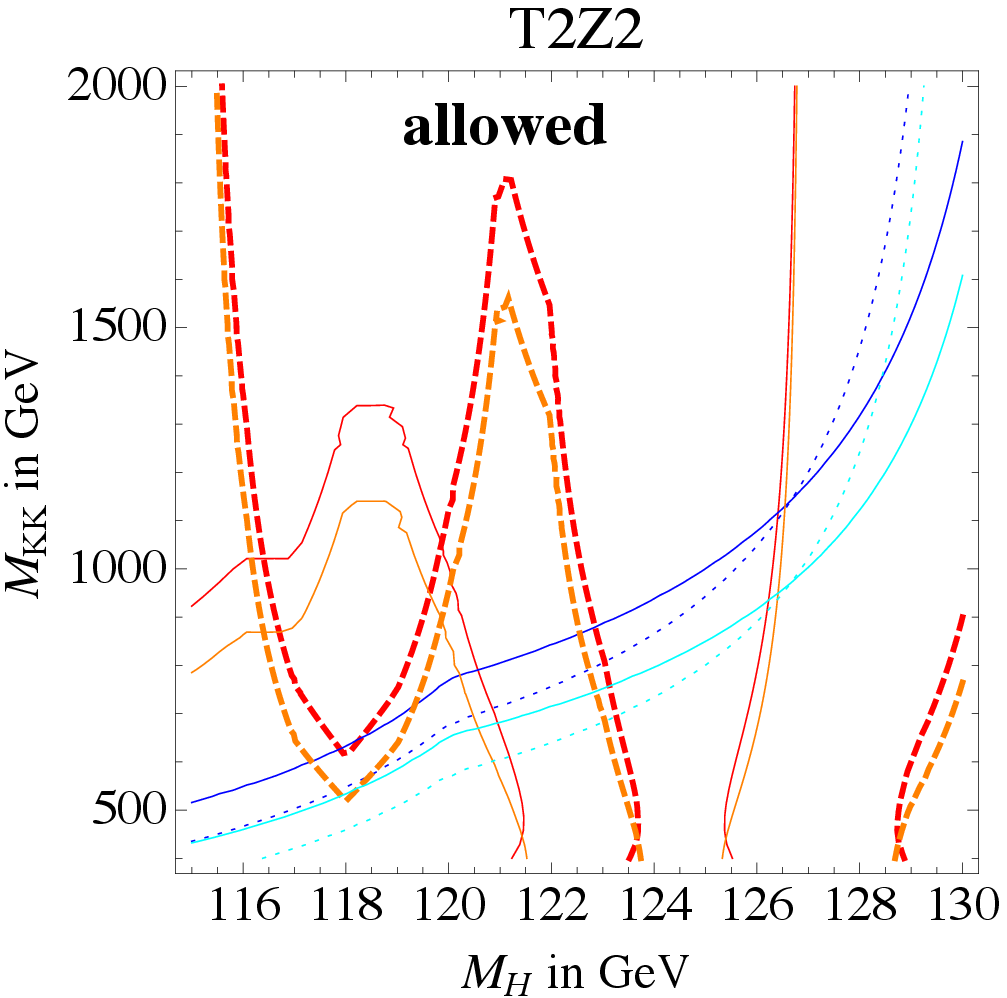}\medskip\\
\includegraphics[width=0.24\columnwidth]{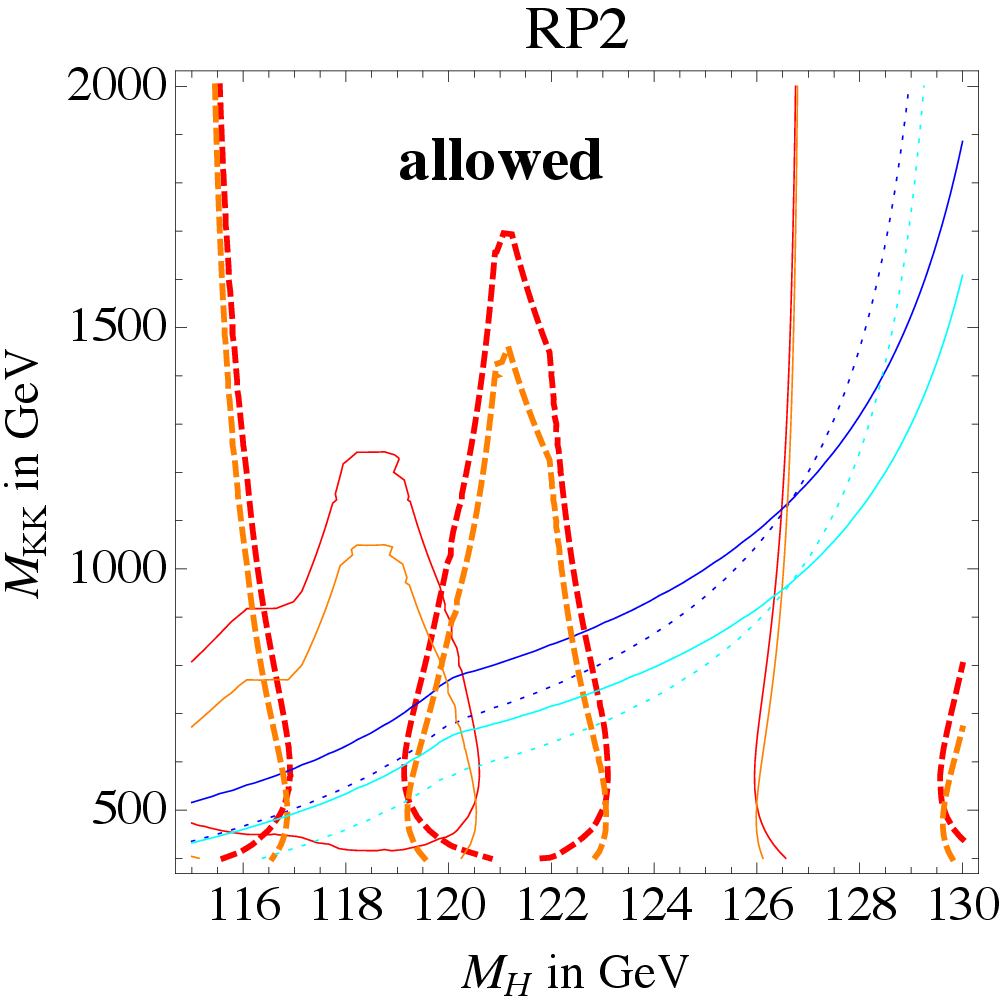}
\includegraphics[width=0.24\columnwidth]{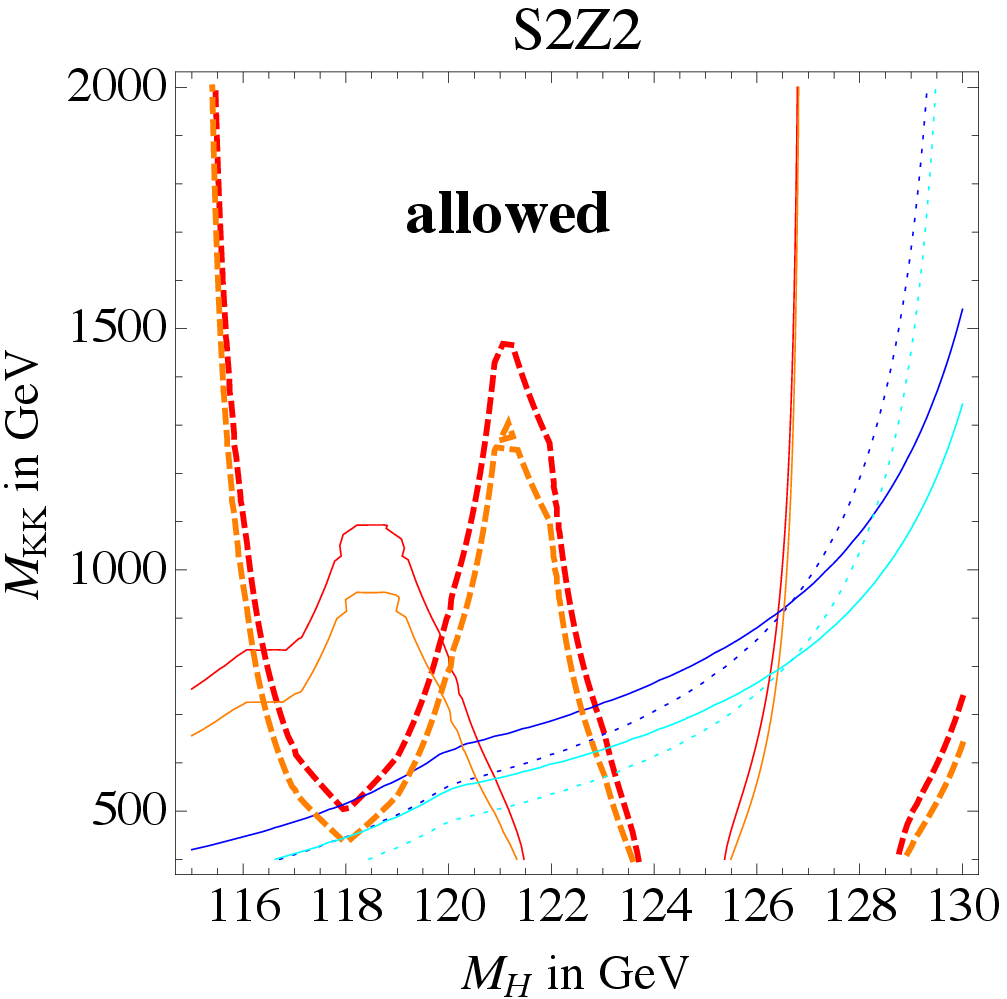}
\includegraphics[width=0.24\columnwidth]{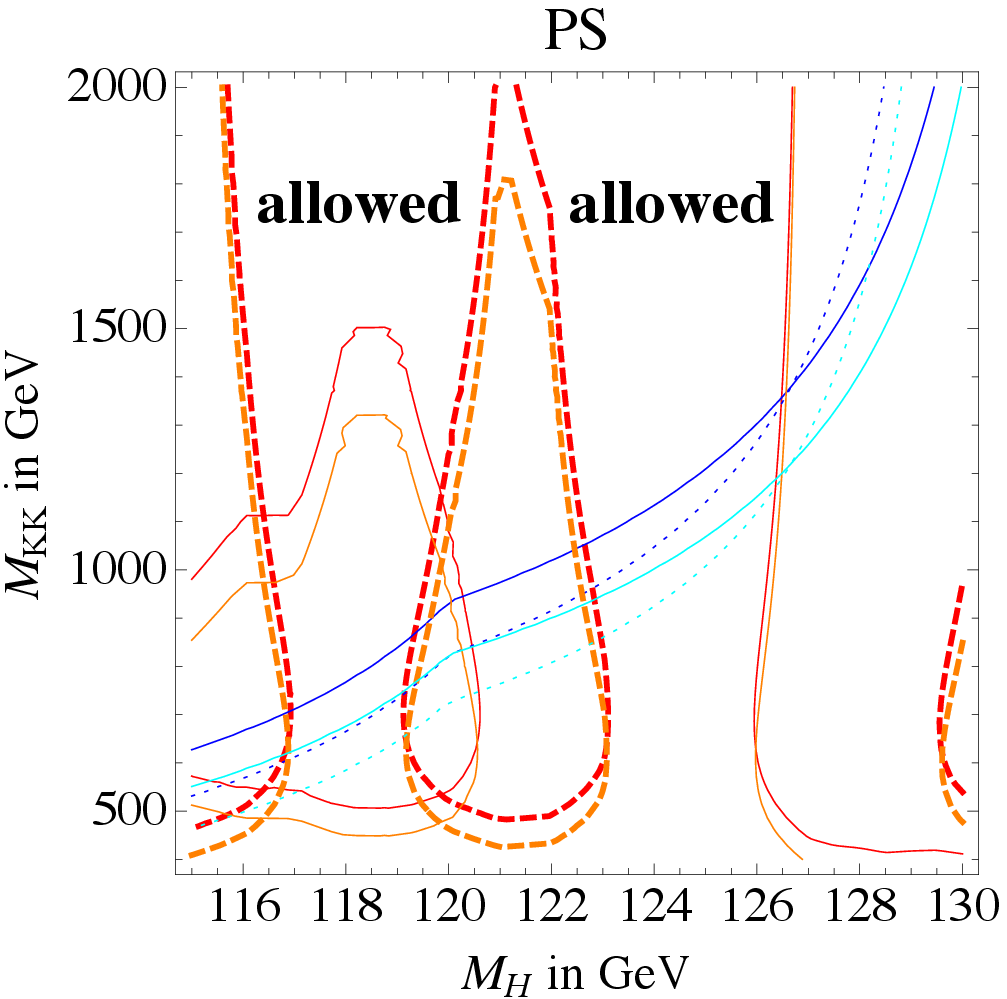}
\includegraphics[width=0.24\columnwidth]{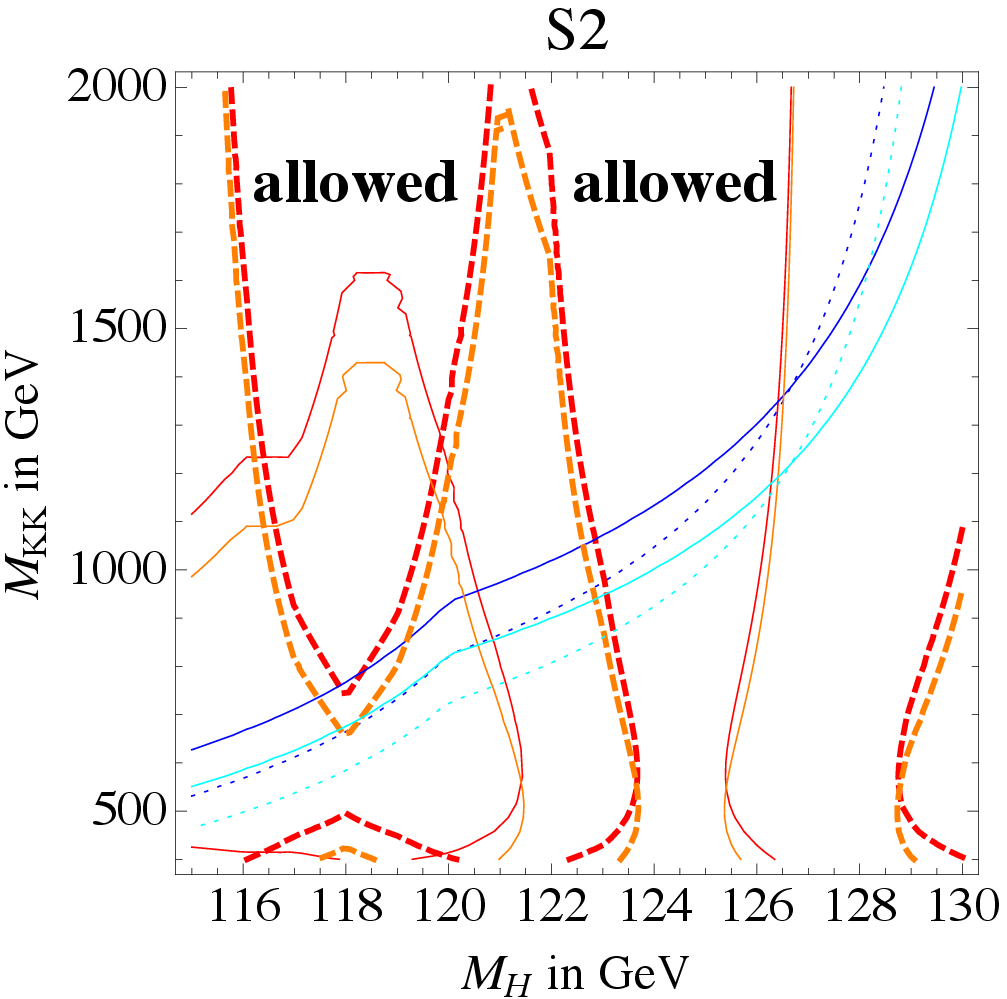}
\caption{95\% CL bounds from $H\to\gamma\gamma$ at ATLAS ({red/orange} dashed) and at CMS (red/orange solid) and from $H\to WW\to l\nu l\nu$ at CMS with cut-based (blue/cyan solid) and with multi-variate BDT (dotted) event selections. The red and blue (orange and cyan) colors correspond to the maximum (minimum) UV cutoff scale in 6D.}\label{fig_bounds}
\end{figure}

\section{LHC bounds on UED models}
In the LHC, the Higgs production is dominated by the gluon fusion process $gg\to H$ induced by the top-quark loop. As a rule of thumb, one can expect that loop-induced UED corrections are significant if a process is prohibited at the tree level in the SM. The gluon fusion is such a process. The KK top quarks make a correction to the Higgs production cross section as
\begin{align}
\hat\sigma_{gg\to H}^\text{UED}
	&=	{\pi^2\over8M_H}\Gamma^\text{UED}_{H\to gg}\,\delta(\hat s-M_H^2),	\\
\Gamma^\text{UED}_{H\to gg}
	&=	K{\alpha_S^2\over8\pi^3}{M_H^3\over v_\text{EW}^2}\left|J_t^\text{SM}+J_t^\text{KK}\right|^2,
\end{align}
where $K$ is the K-factor accounting for the higher order QCD corrections, $\alpha_S$ is the fine structure constant for the QCD, $v_\text{EW}\simeq 246\,\text{GeV}$ is the electroweak scale, and explicit forms of the top and KK-top loop functions $J_t^\text{SM}$ and $J_t^\text{KK}$, respectively, are given in~\cite{Nishiwaki:2011gk,Nishiwaki:2011gm}. As said above, the tree-level widths $\Gamma_{H\to t\bar t}$, $\Gamma_{H\to b\bar b}$, $\Gamma_{H\to c\bar c}$, $\Gamma_{H\to\tau\bar\tau}$, $\Gamma_{H\to WW}$, and $\Gamma_{H\to ZZ}$ are not significantly modified from those in the SM by the KK loop corrections, while the diphoton width becomes
\begin{align}
\Gamma^\text{UED}_{H\to\gamma\gamma}
	&=	{\alpha^2G_FM_H^3\over8\sqrt{2}\pi^3}
			\left|J_W^\text{SM}+J_W^\text{KK}+{4\over3}\left(J_t^\text{SM}+J_t^\text{KK}\right)\right|^2,
\end{align}
where $\alpha$ and $G_F$ are the fine-structure and Fermi constants, respectively, and $J_W^\text{SM}$ ($J_W^\text{KK}$) are loop corrections from {SM-}(KK-) gauge bosons~\cite{Nishiwaki:2011gk}.
{Because of these additional bosonic and fermionic loop correction, Higgs decay to $2\gamma$ {receives} a nontrivial effect.}

The diphoton and $WW$ experimental constraints~\cite{ATLAS_diphoton,CMS_diphoton,CMS_WW} are put on the following ratios, respectively,
\begin{align}
{\sigma^\text{UED}_{gg\to H\to\gamma\gamma}\over\sigma^\text{SM}_{gg\to H\to\gamma\gamma}}
	&\simeq	{\Gamma^\text{UED}_{H\to gg}\Gamma^\text{UED}_{H\to\gamma\gamma}/\Gamma^\text{UED}_{H}
				\over
				\Gamma^\text{SM}_{H\to gg}\Gamma^\text{SM}_{H\to\gamma\gamma}/\Gamma^\text{SM}_{H}},	\label{diphoton_ratio} \\
{\sigma^\text{UED}_{gg\to H\to WW}\over\sigma^\text{SM}_{gg\to H\to WW}}
	&\simeq	{\Gamma^\text{UED}_{H\to gg}/\Gamma^\text{UED}_{H}
				\over
				\Gamma^\text{SM}_{H\to gg}/\Gamma^\text{SM}_{H}},
\end{align}
where we have approximated $\Gamma^\text{UED}_{H\to WW}\simeq\Gamma^\text{SM}_{H\to WW}$ and have taken into account the decay modes into $t\bar t$, $b\bar b$, $c\bar c$, $\tau\bar\tau$, $gg$, $\gamma\gamma$, $W^+W^-$ and $ZZ$ in the total width $\Gamma_{H}$.

In Fig.~\ref{fig_bounds}, we show 95\% CL exclusion plots in $M_\text{KK}$ vs $M_H$ plane from the $H\to\gamma\gamma$ modes at ATLAS~\cite{ATLAS_diphoton} ({red/orange} dashed) and at CMS~\cite{CMS_diphoton} ({red/orange} solid) and from the $H\to WW$ mode at CMS~\cite{CMS_WW} ({blue/cyan}), where solid and dotted lines correspond to the cut-based and BDT event selections for the $WW$ channel, respectively.\footnote{
 As stated above, for all the bounds, we have utilized the values shown in the preliminary version presented on the web. We note that the newer CMS diphoton data set, which we have not utilized, includes vector boson fusion (VBF) events that occurs at the tree level in the SM and hence is not significantly enhanced by the UED loop corrections.}
{The red and blue (orange and cyan) colors correspond to the maximum (minimum) UV cutoff scales in six dimensions; see~\cite{Nishiwaki:2011gk,Nishiwaki:2011gm} for details.}\footnote{We can calculate the processes without UV cutoff dependence in five dimensions.
}
First we can see that the region $115\,\text{GeV}\lesssim M_H\lesssim 127\,\text{GeV}$ is selected by the diphoton exclusion as in the SM. The ATLAS diphoton exclusion around 121\,GeV {became} strong due to { a} statistical fluctuation. In the range $123\,\text{GeV}\lesssim M_H\lesssim 126\,\text{GeV}$, both ATLAS and CMS have an excess of events in the diphoton channel and the bounds from $WW$ signals become stronger. We see that the lower bound for the KK scale is about 500\,GeV--{1}\,TeV depending on the models in this low { Higgs} mass region. The diphoton bounds do not exclude the low KK scale $M_\text{KK}\lesssim 500\,\text{GeV}$ for the lower Higgs mass $M_H\lesssim 123\,\text{GeV}$ in the case of RP2, PS and S2 models, in which we have many low lying KK modes. This is because the KK top contribution $J_t^\text{UED}$ cancels the dominant SM one $J_W^\text{UED}$ in that region.\footnote{
In this parameter region, $J_W^\text{SM}\simeq2$, $J_t^\text{SM}\simeq-0.5$, and $J_W^\text{UED}/J_t^\text{UED}\sim-0.4$.
}
{We can find a similar recent study on mUED in~\cite{Kakizaki_proc}.}

\begin{figure}
\includegraphics[width=0.24\columnwidth]{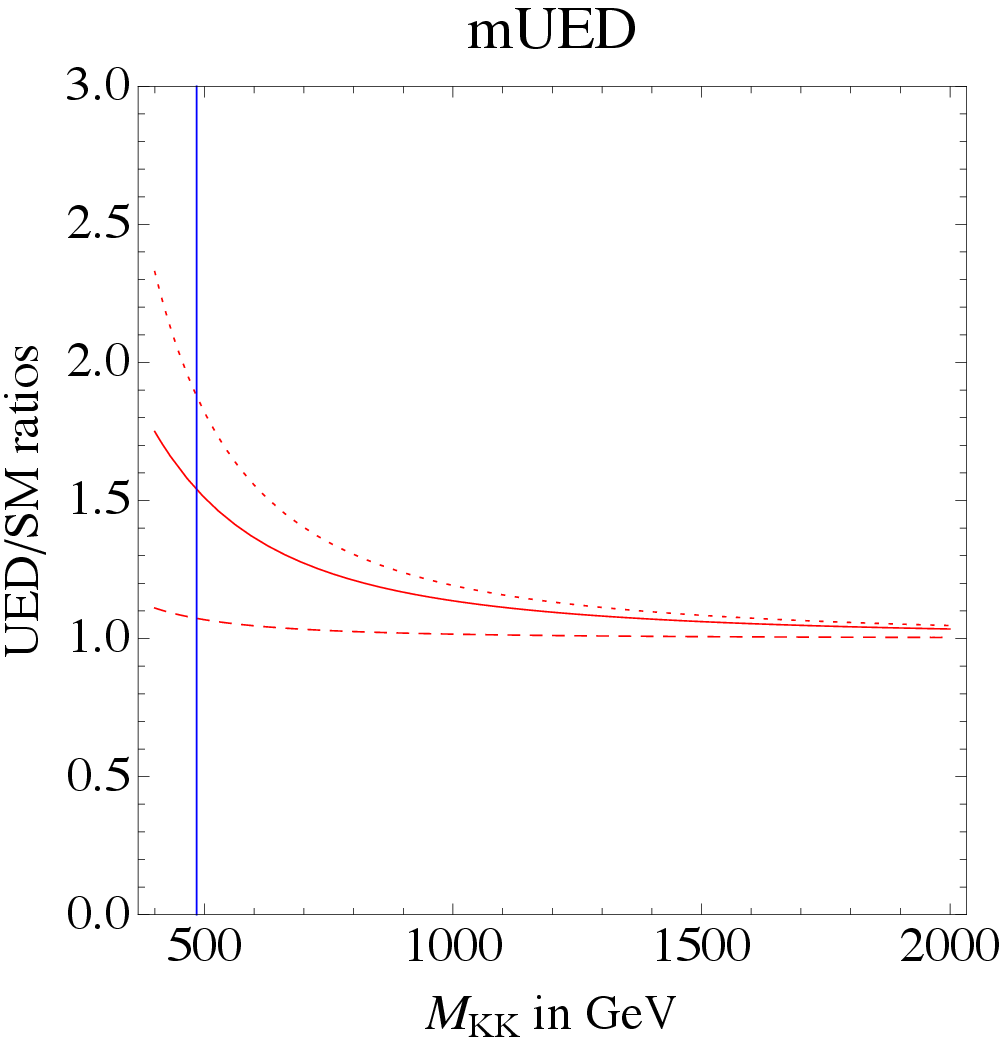}
\includegraphics[width=0.24\columnwidth]{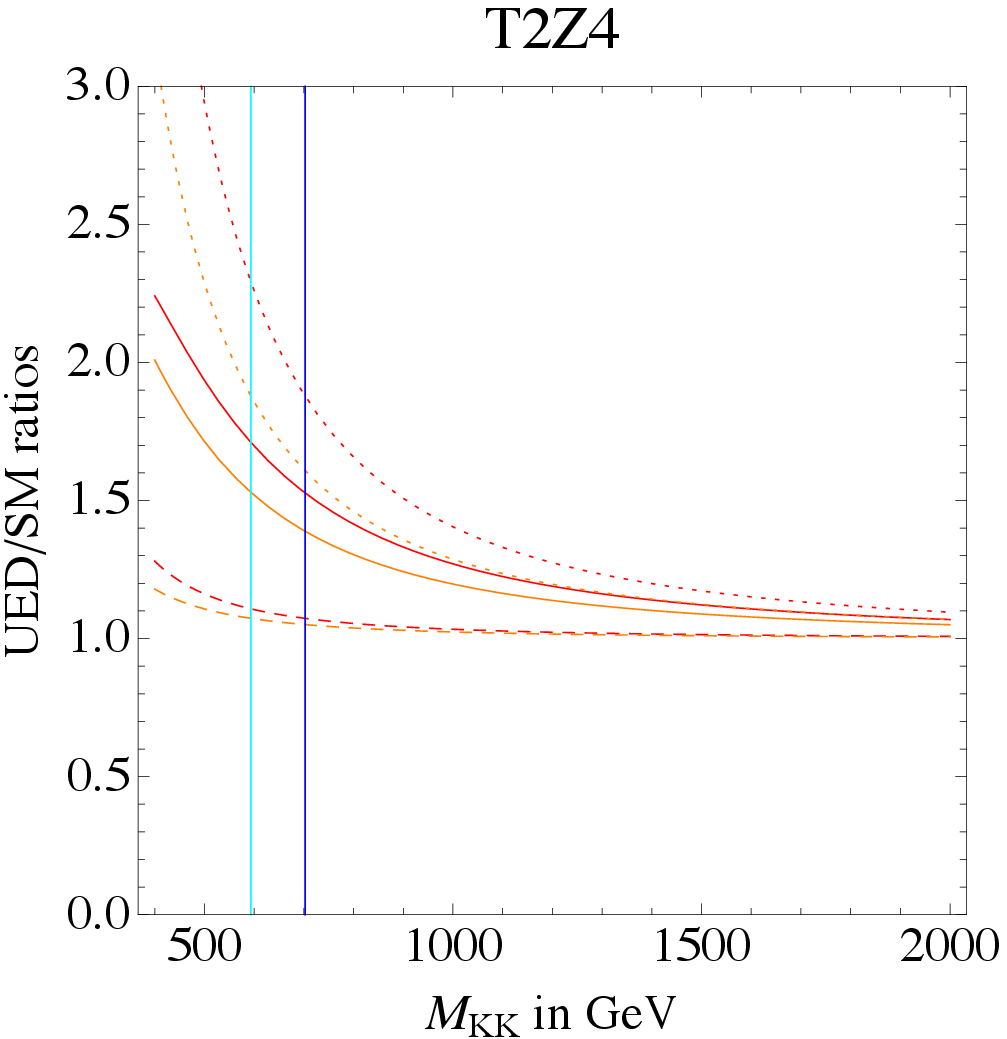}
\includegraphics[width=0.24\columnwidth]{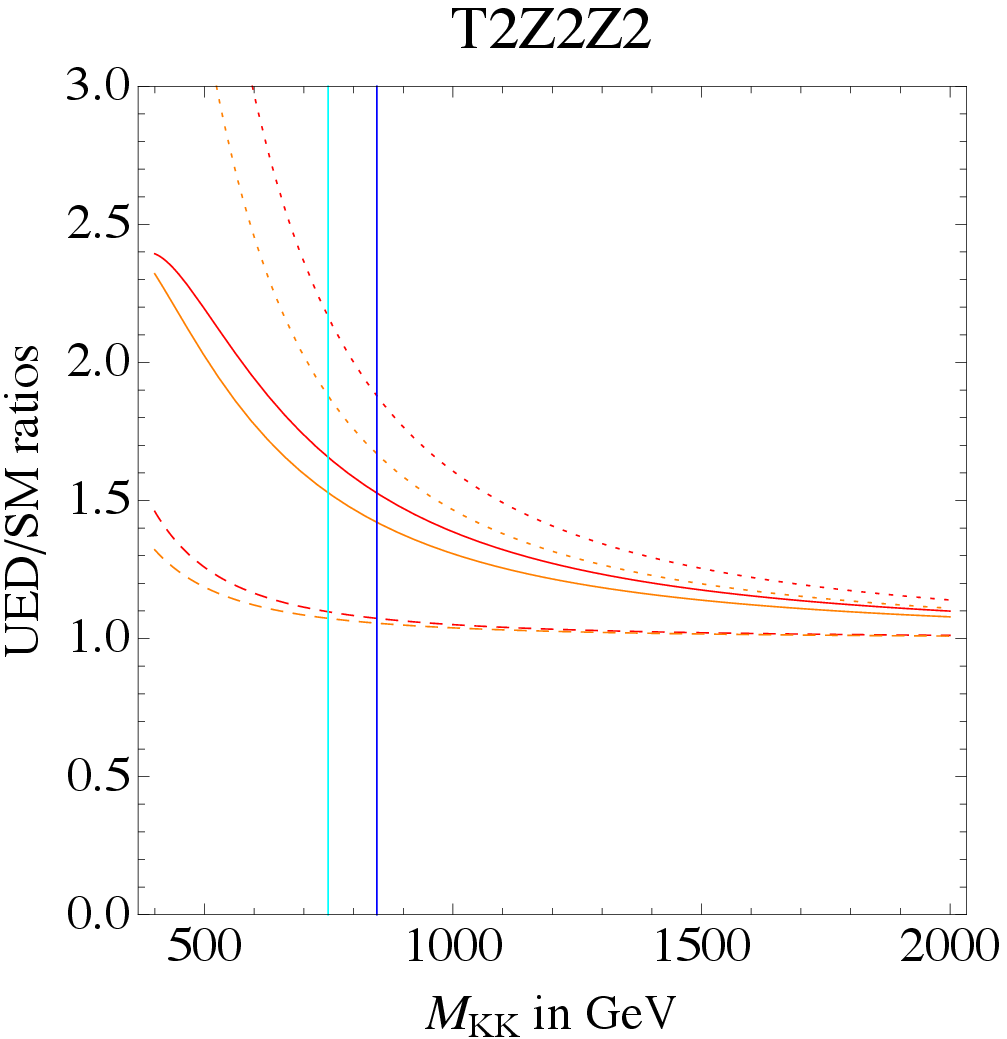}
\includegraphics[width=0.24\columnwidth]{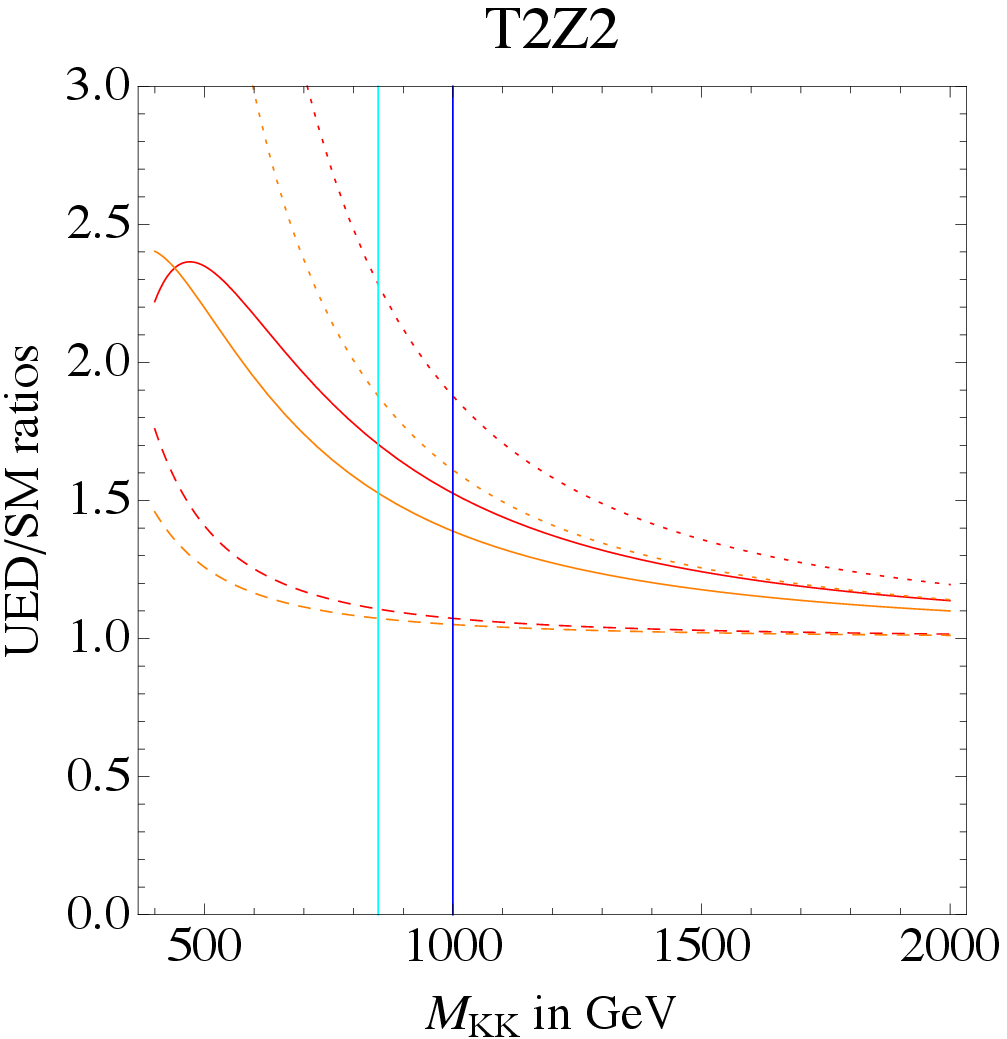}\medskip\\
\includegraphics[width=0.24\columnwidth]{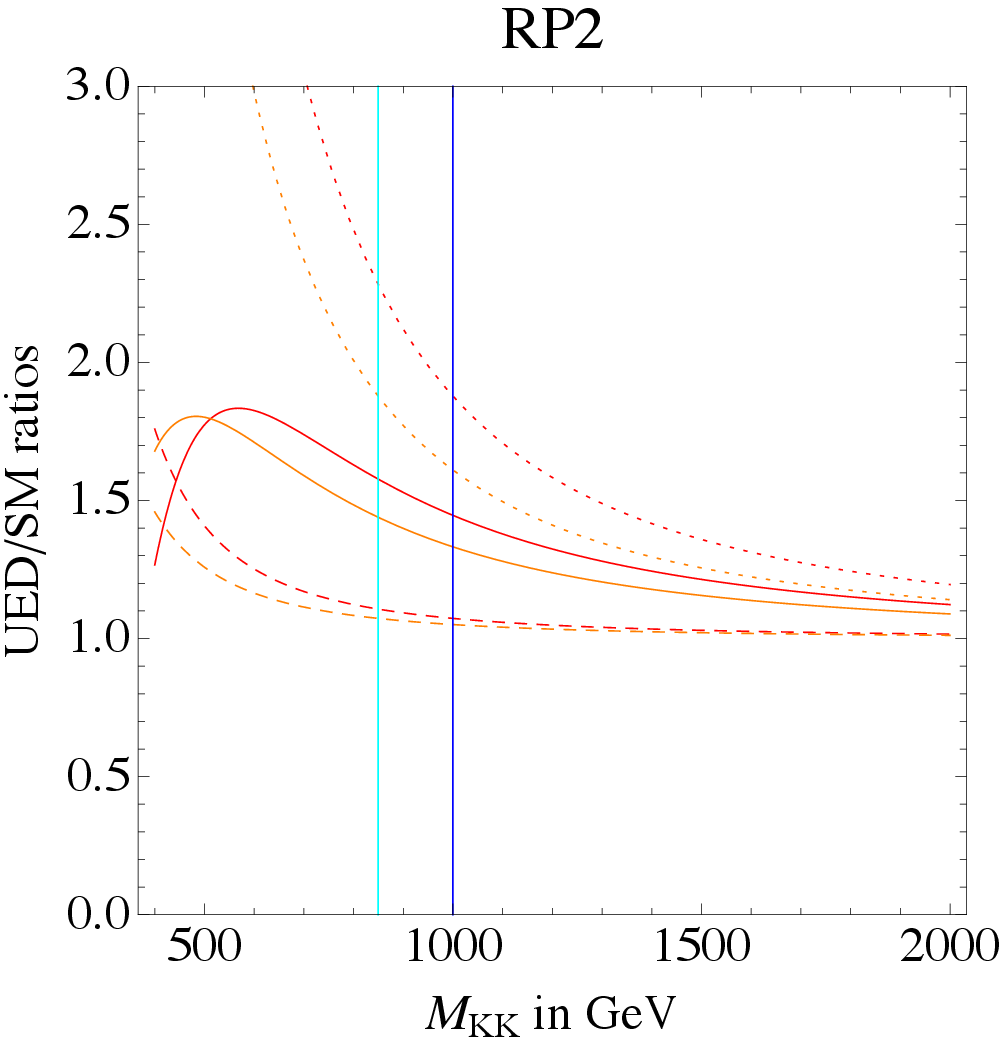}
\includegraphics[width=0.24\columnwidth]{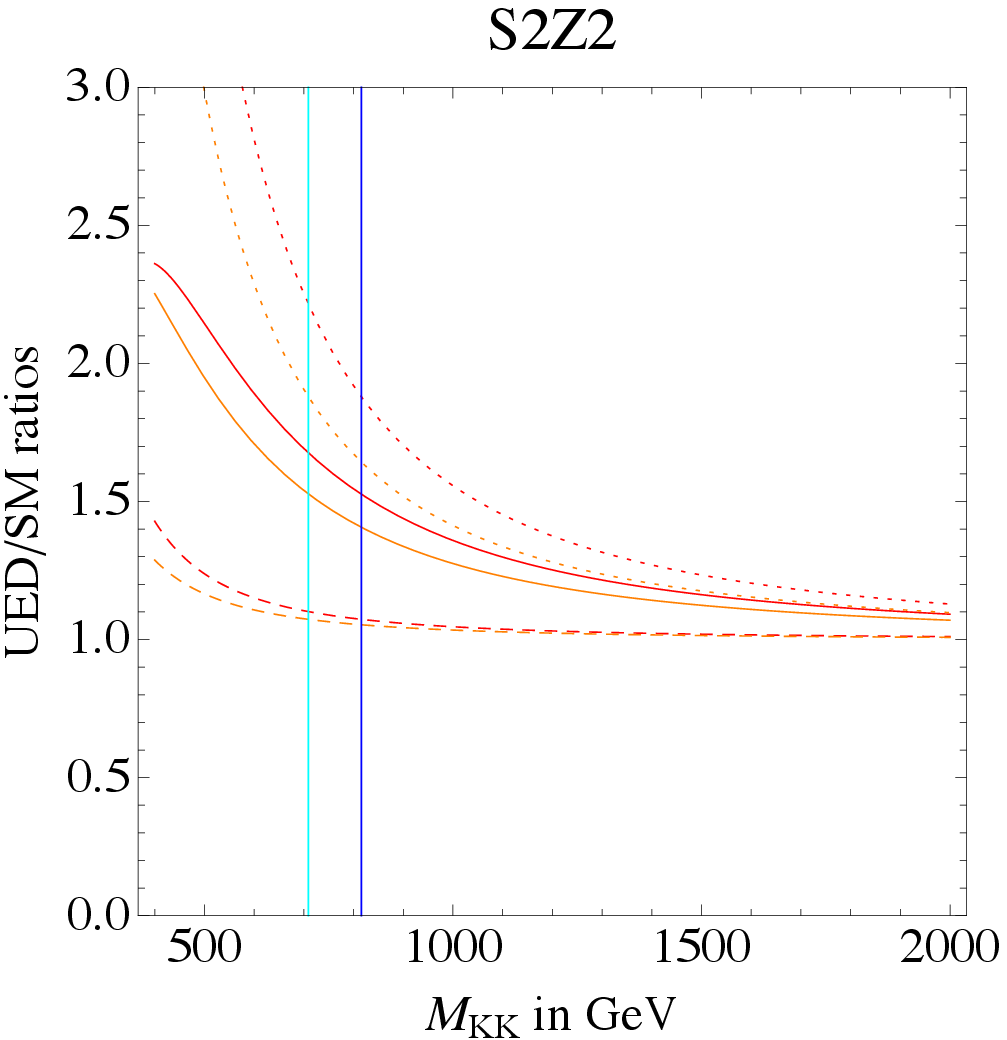}
\includegraphics[width=0.24\columnwidth]{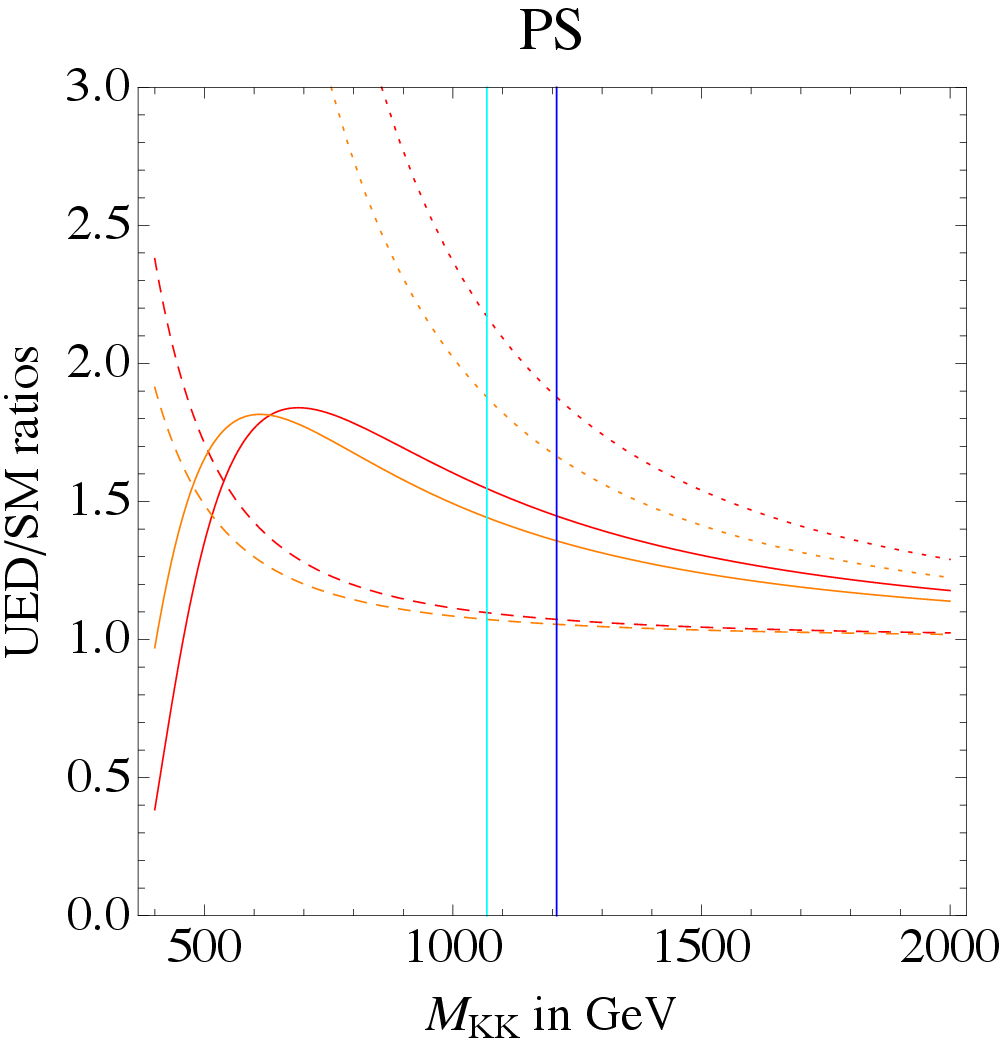}
\includegraphics[width=0.24\columnwidth]{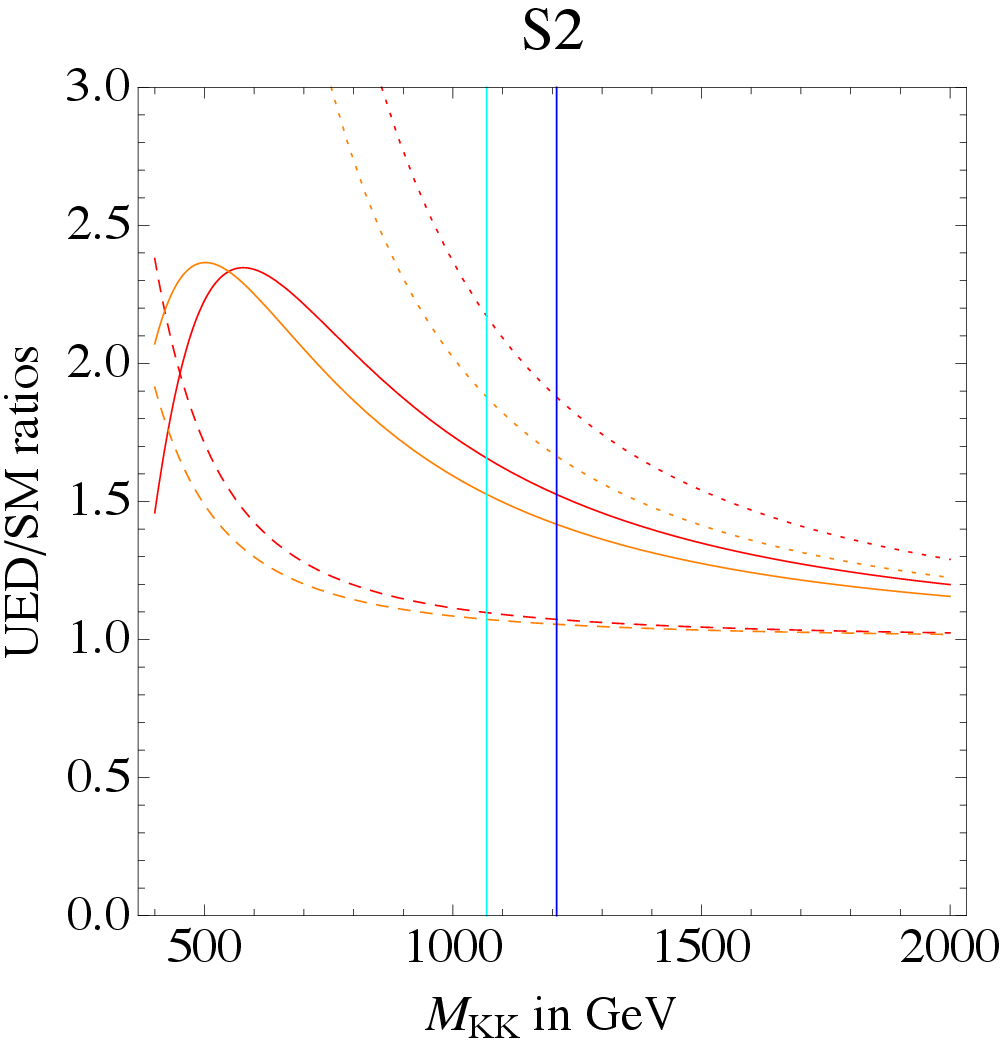}
\caption{Enhancement ratios of UED to SM at $M_H=125\,\text{GeV}$ for the gluon-fusion Higgs production cross section $\sigma_{gg\to H}$ (dotted), for the same with subsequent diphoton decay $\sigma_{gg\to H\to\gamma\gamma}$ (solid), and for the Higgs total decay width $\Gamma_H$ (dashed). The right hand side of the vertical line is allowed by the CMS cut-based $H\to WW$ bound given in Figure~\ref{fig_bounds}. Colors denote the same as in Figure~\ref{fig_bounds}.}\label{fig_enhancements}
\end{figure}

In ATLAS, the best fit value for the ratio of the total Higgs production cross section $\sigma_{gg\to H}/\sigma_{gg\to H}^\text{SM}$ is found to be $\sim1.5$ around the observed excess of events at $M_H\simeq 126\,\text{GeV}$~\cite{ATLAS_combined}. In CMS, the best fit value for the ratio is $\sim0.6$ (1.2) at $M_H=126\,\text{GeV}$ (123--124\,GeV). The preliminary version of Ref.~\cite{ATLAS_combined} reports that the diphoton ratio in Eq.~\eqref{diphoton_ratio} is $\sim2$ at $M_H=126\,\text{GeV}$. Let us examine whether this can be explained by the UED models, keeping in mind the fact that this excess of the cross section ratio is still only $\sim1\sigma$ away from unity.

In Figure~\ref{fig_enhancements}, we plot the enhancement factor for the total Higgs production cross section due to the UED loop corrections (dotted), for the same with subsequent diphoton decay (solid), and also for the total decay width for comparison (dashed) as a function of the first KK mass $M_\text{KK}$. We have chosen $M_H=125\,\text{GeV}$ while the result is insensitive to the Higgs mass in the low mass region $M_H<130\,\text{GeV}$. {Each vertical line shows} the lower bound for the first KK mass $M_\text{KK}$ whose left side is excluded. { Conventions on colors are the same as in Figure~\ref{fig_bounds}.} We see that Higgs cross section {with subsequent diphoton decay} can be enhanced by a factor $\sim1.5$ within the current experimental constraint. Note however that the diphoton ratio (solid) becomes smaller than the $WW$ ratio (dotted) in UED models, in contrast to the observation at ATLAS, where the best fit values for the former and latter are about 2 and 1.2 at the peak. { Note that the $WW$ ratio is almost identical to the ratio for the total production cross section $\sigma_H/\sigma_H^\text{SM}$ (dotted).}

To summarize, the UED corrections become significant for the { SM-}loop induced couplings $Hgg$ and $H\gamma\gamma${; The} {  enhancement of the former} can be seen at LHC, even when multiplied by the {  reduction of the latter} diphoton decay. In the next section, let us see whether the latter reduction can be directly seen at the International Linear Collider (ILC).

\section{ILC and photon photon collider}

\begin{figure}
\includegraphics[width=0.24\columnwidth]{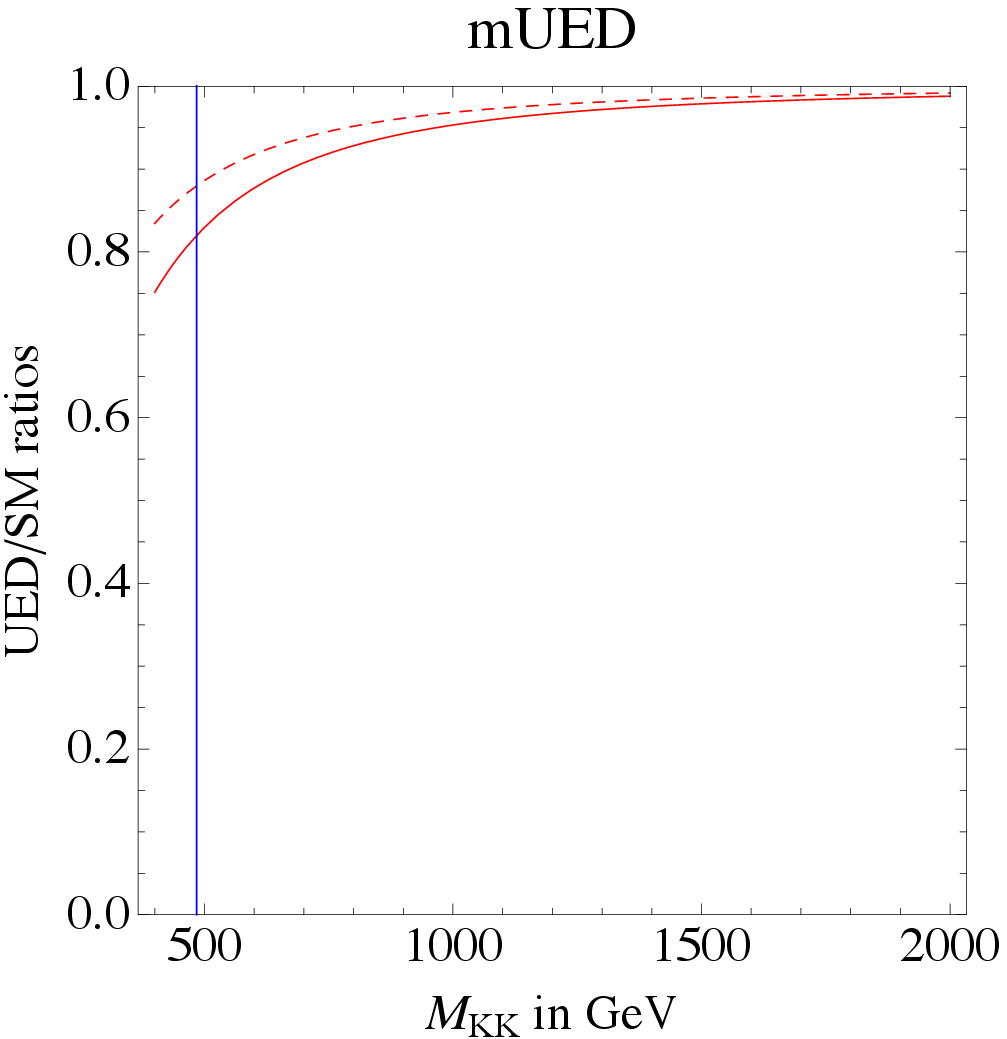}
\includegraphics[width=0.24\columnwidth]{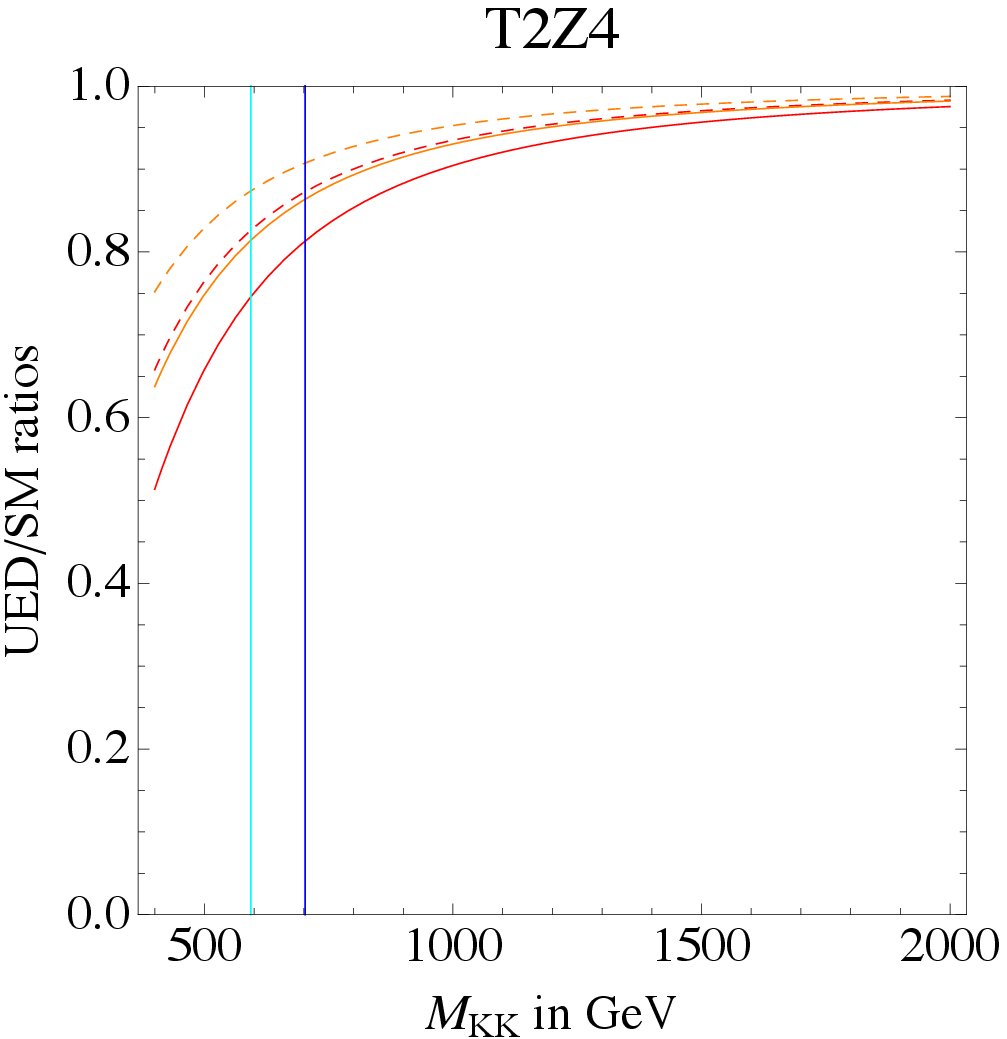}
\includegraphics[width=0.24\columnwidth]{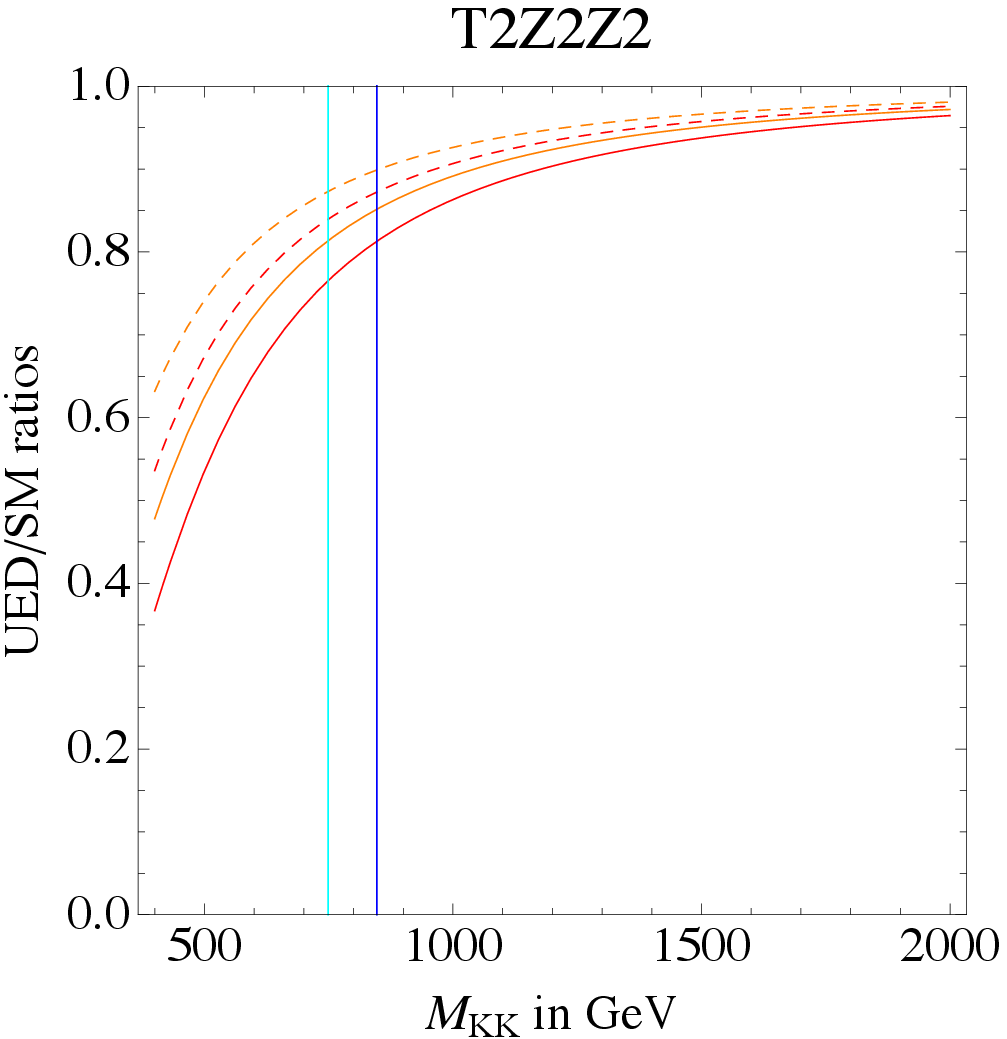}
\includegraphics[width=0.24\columnwidth]{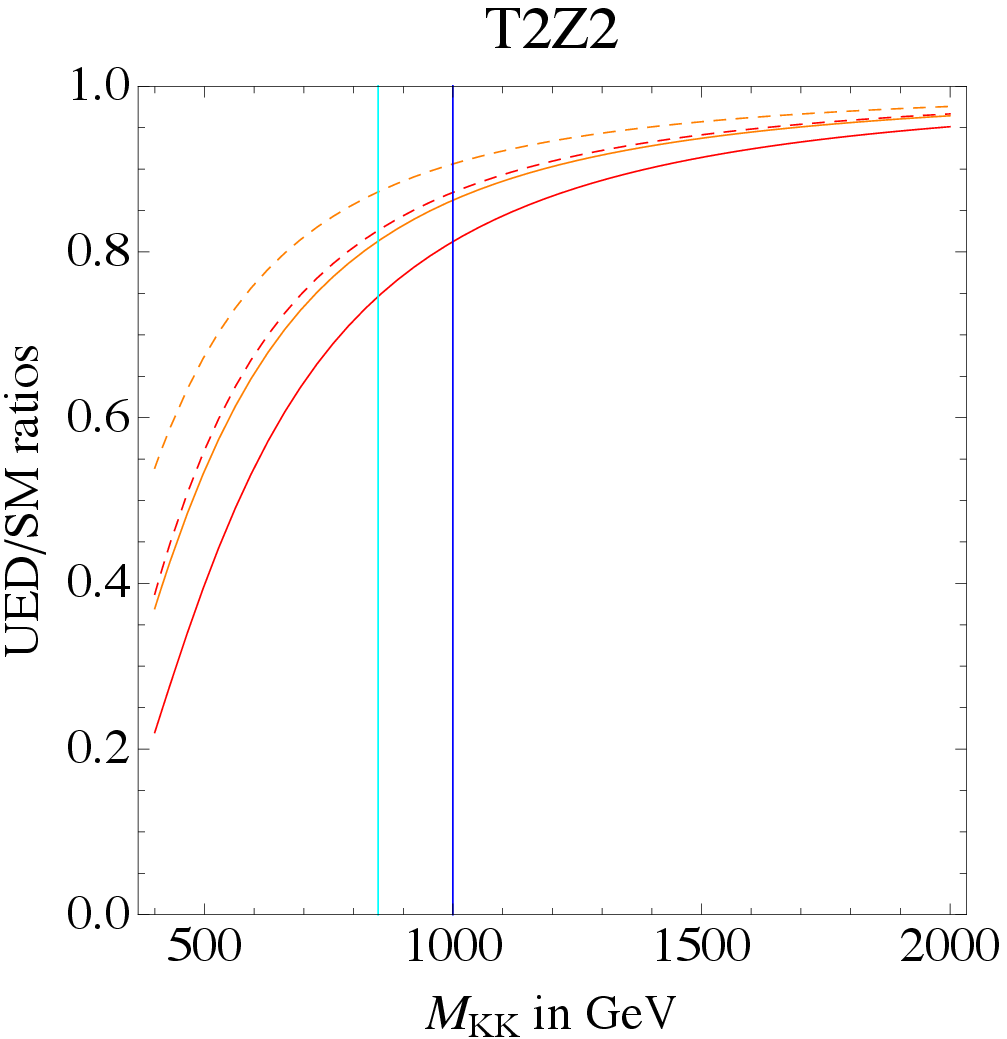}\medskip\\
\includegraphics[width=0.24\columnwidth]{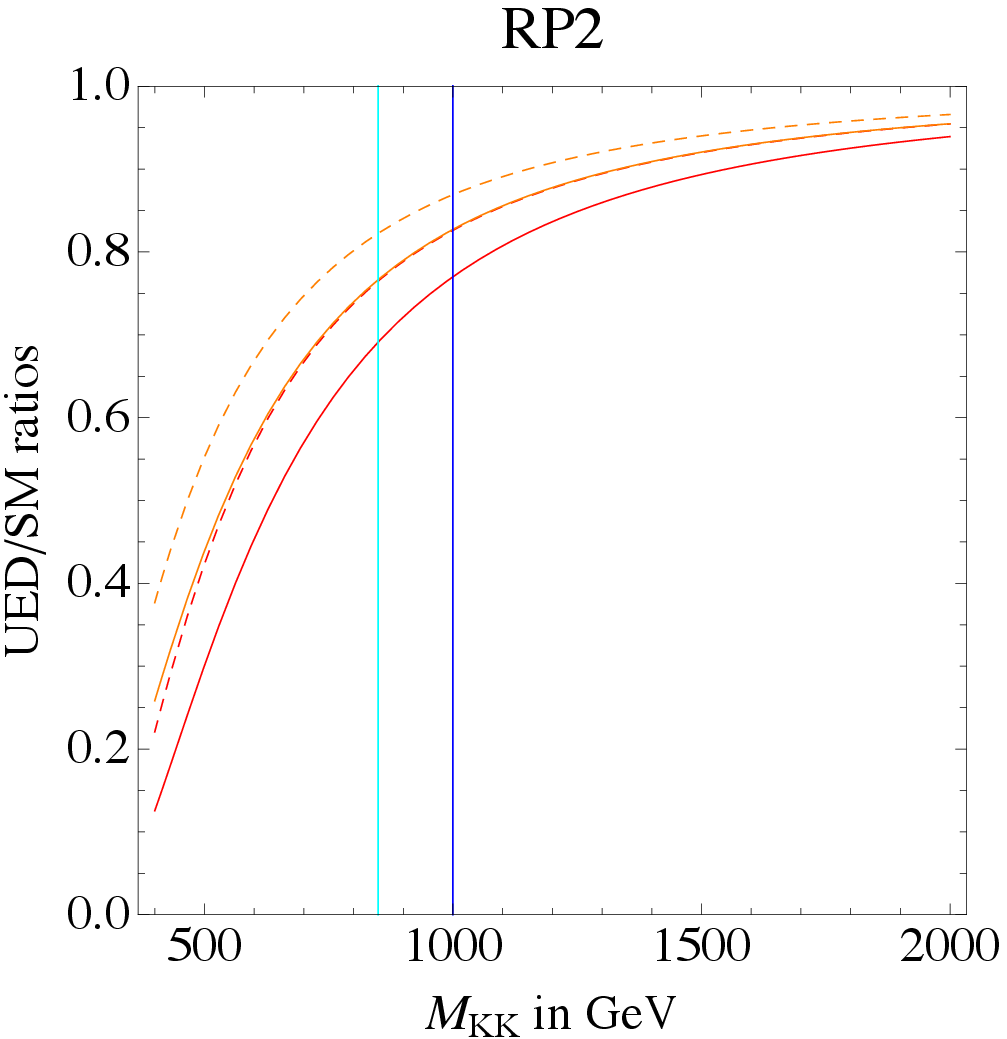}
\includegraphics[width=0.24\columnwidth]{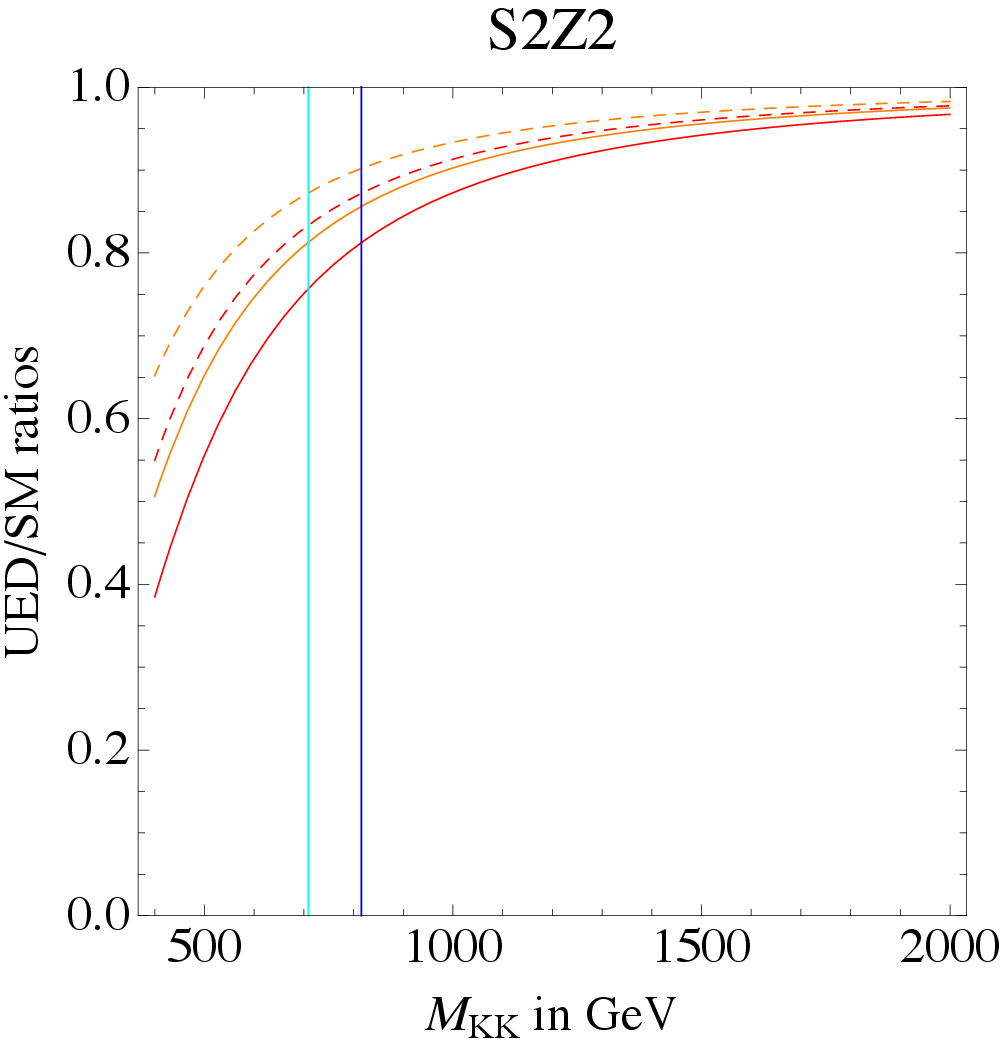}
\includegraphics[width=0.24\columnwidth]{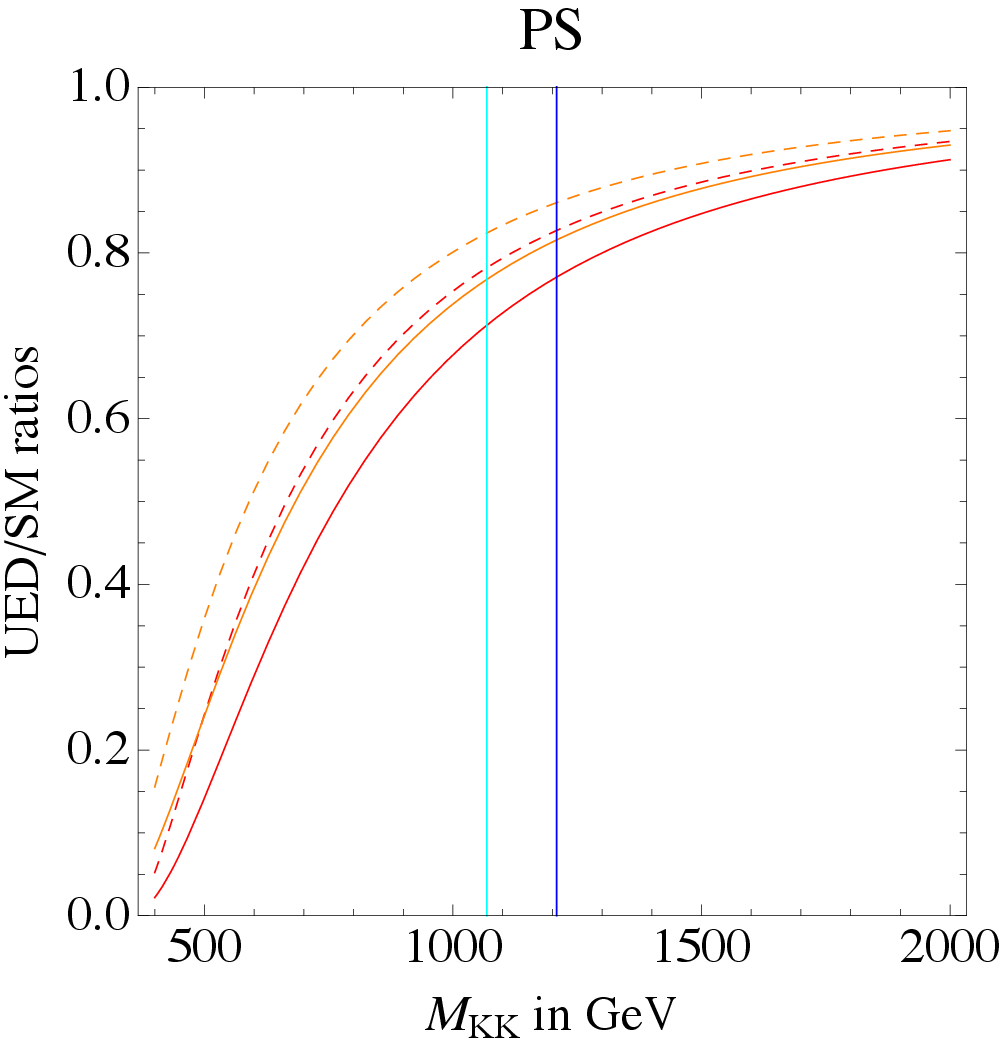}
\includegraphics[width=0.24\columnwidth]{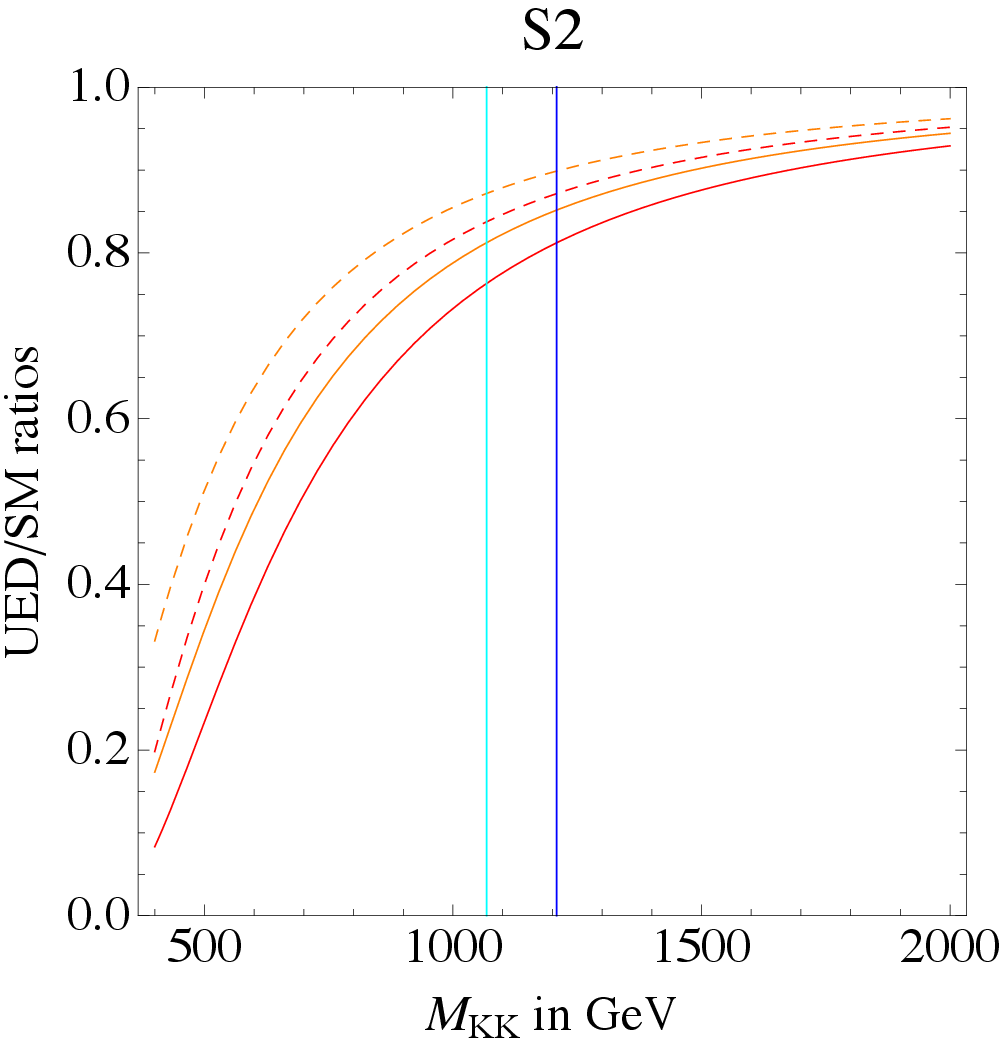}
\caption{Suppression ratios of UED to SM at $M_H=125\,\text{GeV}$ for the Higgs branching ratio into diphoton $\text{BR}(H\to\gamma\gamma)$ (solid) and for the Higgs production cross section at the photon-photon collider $\sigma_{\gamma\gamma\to H}$ (dashed). Colors and vertical lines denote the same as in Figure~\ref{fig_enhancements}.}\label{fig_suppression}
\end{figure}

In Figure~\ref{fig_suppression}, we show the suppression ratio of UED to SM at $M_H=125\,\text{GeV}$ for the Higgs branching ratio { of diphoton decay} $\text{BR}(H\to\gamma\gamma)$ (solid) and for the Higgs production cross section at the photon-photon collider $\sigma_{\gamma\gamma\to H}$ (dashed). Colors indicate the same as in Figure~\ref{fig_enhancements}. The {Higgs decay branching ratio into two photons} is suppressed more than the corresponding decay width because the former is divided by the total decay width that is enhanced by the decay into gluons as shown by the dashed lines in Figure~\ref{fig_enhancements}.

We see that the branching ratio (solid) can be suppressed by a factor $\sim0.8$ within the current experimental bound. This is marginally accessible at the ILC with integrated luminosity $500\,\text{fb}^{-1}$ at $500\,\text{GeV}$ whose expected precision for the $\text{BR}(H\to\gamma\gamma)$ is 23\% for $M_H=120\,\text{GeV}$~\cite{Desch:2003xq}. This precision is refined to 5.4\% with luminosity $1\,\text{ab}^{-1}$ at $1\,\text{TeV}$ for the same Higgs mass~\cite{Barklow:2003hz}.

When we employ the photon photon collider option, $H\gamma\gamma$ coupling can be measured more directly since it becomes the total production cross section of the Higgs. From Figure~\ref{fig_suppression}, we see that the Higgs production cross section (dashed) can be reduced by a factor $\sim 0.9$ in the allowed region to the right of the vertical line. This is well within the reach for an integrated photon-photon luminosity $410\,\text{fb}^{-1}$ at a linear $e^+e^-$ collider operated at $\sqrt{s}=210\,\text{GeV}$ which can measure $\Gamma_{H\to\gamma\gamma} {\times} \text{BR}(H\to b\bar b)$ with an accuracy of 2.1\% for $M_H=120\,\text{GeV}$~\cite{Heinemeyer:2005gs}.

\section{Summary}
In UED models, the loop corrections from the KK-top and KK-gauge bosons modify the $Hgg$ and $H\gamma\gamma$ couplings. Generally we have shown that the former (latter) is enhanced (suppressed) from that in SM, with the former effect dominating the latter. 

We have obtained the 95\% CL allowed region in the $M_\text{KK}$ vs $M_H$ parameter space for all the known UED models in the low mass region $115\,\text{GeV}<M_H<130\,\text{GeV}$ in Figure~\ref{fig_bounds}. 
In this low Higgs mass window, lower and upper bounds for the Higgs mass are given by the ATLAS and CMS diphoton limits, respectively, whereas the lower bound for the KK scale is put by the CMS limit from the $WW\to l\nu l\nu$ channel as $M_\text{KK}\gtrsim 500\,\text{GeV}$--1\,TeV. 

We have also shown the suppression factor from the SM for $\text{BR}(H\to\gamma\gamma)$ and $\Gamma_{H\to\gamma\gamma}$. We see that the former can be suppressed by the factor 0.8 and that this is marginally accessible at the ILC. The $H\gamma\gamma$ coupling itself can also be suppressed by the factor 0.9 which is well within the reach for the photon photon collider option.

\section*{Acknowledgments}
We thank Maria Krawczyk, Shinya Kanemura, and Howard Haber for useful comments in the LCWS11. 


\begin{footnotesize}


\end{footnotesize}



\begin{thebibliography}{99}


\bibitem{ATLAS_combined}
{\it ``Combined search for the Standard Model Higgs boson using up to 4.9\,fb$^{-1}$ of $pp$ collision data at $\sqrt{s}=7\text{TeV}$ with the ATLAS detector at the LHC,''}
Tech.\ Rep.\ ATLAS-CONF-2011-163; arXiv:1202.1408 [hep-ex].



\bibitem{CMS_combined}
S.~Chatrchyan {\it et al.} [CMS Collaboration],
{\it ``Combined results of searches for the Standard Model Higgs boson in $pp$ collisions at $\sqrt{s}=7\text{TeV}$,''}
arXiv:1202.1488 [hep-ex].

\bibitem{Nishiwaki:2011gk}
K.~Nishiwaki, K.~Oda, N.~Okuda and R.~Watanabe,
{\it ``A Bound on Universal Extra Dimension Models from Up to $2\text{fb}^{-1}$ of {LHC} Data At 7{TeV},''}
Phys.\ Lett.\ B {\bf 707} (2012) 506
[arXiv:1108.1764 [hep-ph]].

\bibitem{Nishiwaki:2011gm}
K.~Nishiwaki, K.~Oda, N.~Okuda and R.~Watanabe,
{\it ``Heavy {Higgs} at {Tevatron} and {LHC} in Universal Extra Dimension Models,''}
arXiv:1108.1765 [hep-ph].







\bibitem{ATLAS_diphoton}
The ATLAS Collaboration,
{\it ``Search for the Standard Model Higgs Boson in the diphoton decay channel with 4.9fb$^{-1}$ of $pp$ collisions at $\sqrt{s}=7\text{TeV}$ with ATLAS,''}
arXiv:1202.1414 [hep-ex].



\bibitem{CMS_WW}
The CMS Collaboration, 
{\it ``Search for the Higgs Boson Decaying to W$^+$W$^-$ in the Fully
Leptonic Final State,''}
CMS-PAS-HIG-11-024, (December 2011).

\bibitem{CMS_diphoton}
The CMS Collaboration, 
{\it ``Search for a Higgs boson decaying into two photons in the CMS detector''}
CMS-PAS-HIG-11-030, (December 2011).



{
\bibitem{Kakizaki_proc}
G.~B\'{e}langer {\it et al.},
{\it ``Higgs Phenomenology of Minimal Universal Extra Dimensions,''}
arXiv:1201.5582 [hep-ph].
}



\bibitem{Desch:2003xq}
K.~Desch [Higgs Working Group of the Extended ECFA/DESY Study],
{\it ``Higgs Boson Precision Studies at a Linear Collider,''}
arXiv:hep-ph/0311092.


\bibitem{Barklow:2003hz}
T.~L.~Barklow,
{\it ``Higgs Coupling Measurements at a 1-Tev Linear Collider,''}
arXiv:hep-ph/0312268.


\bibitem{Heinemeyer:2005gs}
S.~Heinemeyer {\it et al.},
{\it ``Toward High Precision Higgs-Boson Measurements at the International Linear $e^+e^-$ Collider,''}
arXiv:hep-ph/0511332.




\end{thebibliography}
\end{document}